\documentclass[10pt,journal,compsoc]{IEEEtran}
\usepackage{mathptmx}
\usepackage{graphicx}
\usepackage{times}
\usepackage{amssymb}
\usepackage{pifont}
\newcommand{\cmark}{\ding{51}}
\newcommand{\xmark}{\ding{55}}

\usepackage[bookmarks,backref=true,linkcolor=black]{hyperref} 
\hypersetup{
  pdfauthor = {},
  pdftitle = {},
  pdfsubject = {},
  pdfkeywords = {},
  colorlinks=true,
  linkcolor= black,
  citecolor= black,
  pageanchor=true,
  urlcolor = black,
  plainpages = false,
  linktocpage
}

\usepackage{textcomp}
\usepackage{color,soul}
\usepackage[dvipsnames]{xcolor}
\usepackage{todonotes}
\usepackage{soul}
\usepackage{wrapfig}
\usepackage[final]{changes}

\usepackage[
singlelinecheck=false 
, font={footnotesize,sf}
]{caption}

\usepackage{enumitem}

\usepackage[english]{babel}
\usepackage{hyperref}
\addto\extrasenglish{%

}

\definecolor{cb_orange}{rgb}{1.0,0.51,0.0}
\definecolor{cb_blue}{rgb}{0.22,0.49,0.72}
\definecolor{cb_green}{rgb}{0.3,0.67,0.29}
\definecolor{cb_red}{rgb}{0.89,0.1,0.11}
\definecolor{cb_purple}{rgb}{0.6, 0.31, 0.64}
\definecolor{cb_brown}{rgb}{0.6, 0.4, 0.2}
\definecolor{cb_crimson}{rgb}{0.86, 0.08, 0.24}

\begin{document}

\def\lsituated{\textsc{Situated}}
\def\lbound{\textsc{Boundary}}
\def\lboundary{\textsc{Boundary}}
\def\lheight{\textsc{Height}}
\def\langle{\textsc{Angle}}
\def\lvalue{\textsc{Value}}

\def\tidentify{\textsc{Identify}}
\def\tcompare{\textsc{Compare}}
\def\tsummarize{\textsc{Summarize}}
%
\title{Labeling Out-of-View Objects in Immersive Analytics to Support Situated Visual Searching}
%
%
%
%


\author{Tica~Lin, 
        Yalong~Yang, 
        Johanna~Beyer 
        and~Hanspeter~Pfister 
\IEEEcompsocitemizethanks{\IEEEcompsocthanksitem Tica Lin,  Johanna Beyer, and Hanspeter Pfister are with John A. Paulson School of Engineering and Applied Sciences, Harvard University, Cambridge, MA, 02138.
E-mail: \{mlin, jbeyer, pfister\}@g.harvard.edu
\IEEEcompsocthanksitem Yalong Yang is with the Department of Computer Science,  Virginia Tech, Blacksburg, VA, 24060. E-mail: yalongyang@vt.edu
}
\thanks{Manuscript received April 19, 2005; revised August 26, 2015.}}

%
%

\markboth{Journal of \LaTeX\ Class Files,~Vol.~14, No.~8, August~2015}%
{Shell \MakeLowercase{\textit{et al.}}: Bare Demo of IEEEtran.cls for Computer Society Journals}
%



\IEEEtitleabstractindextext{%

\begin{abstract}
Augmented Reality (AR) embeds digital information into objects of the physical world. Data can be shown \emph{in-situ}, thereby enabling real-time visual comparisons and object search in real-life user tasks, such as comparing products and looking up scores in a sports game.  
While there have been studies on designing AR interfaces for situated information retrieval, there has only been limited research on AR object labeling for visual search tasks in the spatial environment.
In this paper, we identify and categorize different design aspects in AR label design and report on a formal user study on labels for out-of-view objects to support visual search tasks in AR. 
We design three visualization techniques for out-of-view object labeling in AR, which respectively encode the relative physical position (height-encoded), the rotational direction (angle-encoded), and the label values (value-encoded) of the objects. 
We further implement two traditional in-view object labeling techniques, where labels are placed either next to the respective objects (situated) or at the edge of the AR FoV (boundary). 
We evaluate these five different label conditions in three visual search tasks for static objects.
Our study shows that out-of-view object labels are beneficial when searching for objects outside the FoV, spatial orientation, and when comparing multiple spatially sparse objects.  
Angle-encoded labels with directional cues of the surrounding objects have the overall best performance with the highest user satisfaction.
We discuss the implications of our findings for future immersive AR interface design. 
\end{abstract}

}

\maketitle

\IEEEdisplaynontitleabstractindextext

%
\IEEEpeerreviewmaketitle

\IEEEraisesectionheading{\section{Introduction}\label{sec:introduction}}

%
%
%
%

\begin{figure*}[t!]
    \centering
    \includegraphics[width=\linewidth]{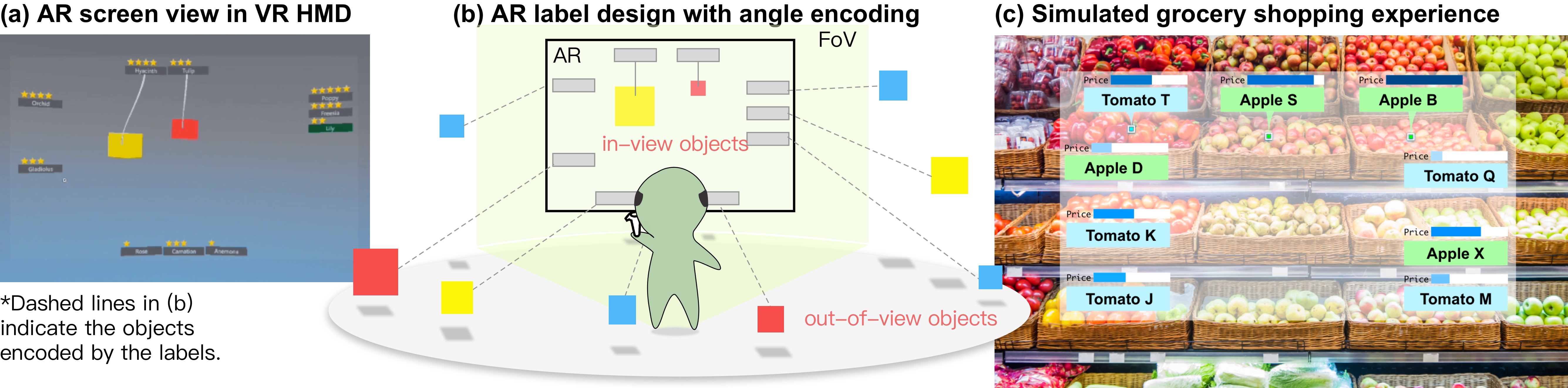}
    \caption{AR label design for objects outside the field-of-view (FOV).
     (a) Our user study uses a VR HMD to simulate consistent AR conditions.
    (b) The user is surrounded by spatially sparse objects and can see labels for in-view and out-of-view objects on the AR screen.
     Labels for out-of-view objects are placed on the boundary to support embodied navigation. 
     (c) A simulated grocery shopping experience with AR labels.
    } 
    \label{fig:teaser}
\end{figure*}

\IEEEPARstart{S}{earching} 
for elements of interest is an essential component for almost all visual analytics tasks, such as \emph{identifying} outliers and clusters, \emph{comparing} values encoded in visual elements, and \emph{summarizing} properties for a group of objects.
Designers of desktop visualization tools have grappled with this problem for decades and have developed sophisticated techniques to support visual search. 
To make search results pop out in a visualization, it is common to show icons or labels overlaid on top of the original visualization (e.g., restaurants on a map).
To avoid occluding local visualization features, external labeling techniques~\cite{bekosExternalLabelingTechniques2019} place labels outside the visualization, and use leader lines to connect the visual elements with their associated labels. 
Other techniques allow users to zoom out to an overview to locate the searched for elements, and then zoom in to check the details.
These techniques are well studied and widely used in desktop visualization tools.

With the rapid development in technology, consumer-level head-mounted displays (HMDs) for virtual reality (VR) and augmented reality (AR) have become mature and affordable. 
In response to this revolution, immersive analytics has emerged as a new research field, focusing on the use of engaging, embodied analysis tools in VR and AR, to support data understanding and decision making~\cite{marriott_situated_2018}.
VR/AR removes the boundaries of physical screens and facilitates new human-computer interaction experiences.
In particular, AR can embed digital information in the physical world in almost any environment. Thereby, AR provides opportunities to support situated visual search for real-life tasks~\cite{ens_grand_2021},
such as 
showing ratings of nearby restaurants~\cite{willettEmbeddedDataRepresentations2017}, 
adding historical information for artifacts in a museum~\cite{miyashita2008augmented}, 
or annotating player names and stats in a basketball game~\cite{lin2020sportsxr,courtvision}, all without diverging users' attention from the real-world environment. 

In these AR applications, additional information for the physical objects is commonly rendered in small ``virtual canvases'' or labels. 
However, compared to the desktop, visual searching in AR has two distinct differences.
First, interface designers cannot manipulate everything in the AR display space.
While labels can be moved freely in AR, physical objects (e.g., the locations of the restaurants, the basketball players on the court) usually cannot be manipulated. As a result, users have to physically move or rotate to get to or face the physical targets.
Second, unlike desktop displays where visual elements can be scaled to fit the screen, physical objects in the real world cannot be scaled (e.g., again, a restaurant or a basketball player).
Due to these two distinct differences, some desktop techniques for visual searching, such as zooming, cannot be used in AR applications, and more emphasis needs to be put on efficient object labeling techniques.
Therefore, further studies are needed to understand \textbf{how to place labels in AR to effectively support situated visual search}. Currently, there are still many open questions, such as whether labels should be placed as close to their linked objects as possible, or whether labels should also encode spatial information of objects that are inside or outside the user's FOV to help reduce a user's search effort.

To address these questions, we first thoroughly review existing labeling techniques in AR.
Previous studies on labeling objects in AR focused on algorithms to place labels~\cite{bellViewManagementVirtual2001,maassDynamicAnnotationInteractive2006,azumaEvaluatingLabelPlacement2003, tatzgernHedgehogLabelingView2014,grassetImagedrivenViewManagement2012} and filter data~\cite{tatzgernAdaptiveInformationDensity2016} to avoid visual clutter on limited AR FoV.
Most labeling techniques only show labels for objects that are within the user's FoV, so-called in-view objects. 
In many real-world AR scenarios, however, the objects of interest are not just in front of the user. Sometimes the user may not even know the locations of the target objects.
Without any additional clues, the user will then have to physically scan the entire surrounding environment to find an object and its label, which is a time-consuming and fatiguing task.
To support efficient situated visual search tasks, labels for out-of-view objects (i.e., physical objects that are not within the FoV) is thus necessary.
Although previous work studied visual encoding techniques for out-of-view objects in AR~\cite{schinkeVisualizationOffscreenObjects2010,jo2011aroundplot,siuSidebARsImprovingAwareness2013,gruenefeldEyeSee360DesigningVisualization2017,borkEfficientVisualGuidance2018}, not much research has focused on labels and label design for out-of-view objects. In this work, we explore different visual encodings for AR labels and quantitatively compare them for situated analysis tasks in a user study.

The motivation for this paper is to investigate in which situations out-of-view labels are helpful, and whether certain tasks would benefit from only providing in-view labels.
We specifically focus on labeling static objects as a preliminary step towards answering this question. Labeling static objects has many real-world use scenarios, such as goods in grocery stores or restaurants around a person (see \autoref{sec:usage_scenario}). Labeling moving objects, on the other hand, requires other design considerations, which we briefly discuss in our future work (\autoref{sec:future}).

Our first contribution is \textbf{a systematical exploration of the design space of labels in AR applications, with an emphasis on labeling out-of-view objects}.
We analyzed previous works on AR labeling~\cite{bellViewManagementVirtual2001,grassetImagedrivenViewManagement2012}, embodied interaction~\cite{satriadiAugmentedRealityMap2019}, and leader line placement~\cite{tatzgernHedgehogLabelingView2014}, and identified four different aspects in AR labeling systems (label proxy, proximity, encoding, and interaction), which can guide the design of future AR labeling systems.

Our second contribution is \textbf{a controlled user study comparing five representative AR labeling techniques}.
In two conditions we show labels for in-view objects only:
\emph{situated}, where labels are placed slightly above the physical object, and
\emph{boundary}, where labels are pushed to the boundaries of the FoV to reduce visual clutter.
We also designed three conditions for labeling of out-of-view objects:
\emph{height}, where labels are placed at the same height as their linked objects will appear once the user rotates toward them;
\emph{angle}, where label positions indicate the angle between the user's viewing direction and the object, using a top-down view metaphor; and
\emph{value}, where labels are ordered by their associated values (e.g., restaurant ratings) on the left of the FoV.
Following previous work in AR research~\cite{marquardtComparingNonVisualVisual2020,raganSimulationAugmentedReality2009,jungEnsuringSafetyAugmented2018}, to easily manipulate the locations of physical objects, we used VR to simulate AR applications in our study. 
To achieve this, we created an AR FoV in VR, and render labels and leader lines only when they are within this AR FoV.
In our study, we limit the number of labels to 10 and 20 objects.
This is based on the number of AR labels used in other user studies~\cite{tatzgernHedgehogLabelingView2014,madsenTemporalCoherenceStrategies2016}, and also follows the design guideline of avoiding information overload in AR applications~\cite{feinerKnowledgebasedAugmentedReality1993,julierInformationFilteringMobile2002,tatzgernAdaptiveInformationDensity2016}. 
For labeling a larger number of objects, label clustering and filtering techniques have been proposed~\cite{tatzgernAdaptiveInformationDensity2016}, but this is outside the scope of our study.

We evaluated our five different label conditions in three visual search tasks. 
Overall, we found that labels for out-of-view objects helped users in getting a quick overview, provided helpful directional cues, and was beneficial for visual comparisons in AR.
We found that the \emph{angle} condition was the overall winner; it was faster than \emph{value} in all tested tasks, faster than \emph{situated} and \emph{boundary} labels in value comparisons, and faster than \emph{height} in summarizing multiple data points.
The \emph{angle} condition was also preferred by our participants overall.
Our work contributes to the growing body of knowledge of 
effective AR interface design to support situated analysis in the real world.

\section{Related Work}


In the following, we review the literature on situated visual search to motivate the use case of AR labels, and point out the gap in previous AR labeling and visualization techniques.


\subsection{Situated Visual Search in AR}
\label{sec:situated_visual_searching}
Visual search is a fundamental perceptual task that people perform to identify a particular target among other distractors in a visual environment~\cite{wolfeVisualSearchScenes2011}. 
In real-world visual search tasks, information relevant to the location or object can be placed \textit{situated} to support target comparison and spatial navigation~\cite{bioccaAttentionFunnelOmnidirectional2006}. 
Many design frameworks and use cases have been proposed that exploit the situatedness~\cite{willettEmbeddedDataRepresentations2017,elsayed_situated_2015} and embodiment~\cite{satriadiAugmentedRealityMap2019} of AR visualizations.
Buschel et al.~\cite{buschelHereNowRealityBased2018} proposed a conceptual framework for annotating real-world objects with situated labels and accessing information through interaction with the labels. Bach et al.~\cite{bachDrawingARCANVASDesigning} proposed AR-canvas, a framework for designing embedded visualizations for visual search in realistic use scenarios.
Attention management research~\cite{horvitzModelsAttentionComputing2003, vertegaalDesigningAttentiveInterfaces} has shown that high information density can lead to information overload, especially in real-world use cases~\cite{redelmeierAssociationCellularTelephoneCalls1997,pascoal2017information}.
Managing information density on AR screens can be achieved with spatial and knowledge-based filtering~\cite{julierInformationFilteringMobile2002,feinerKnowledgebasedAugmentedReality1993}, and data clustering~\cite{tatzgernAdaptiveInformationDensity2016}. 
Building on the existing AR information retrieving workflow, our study is a concrete step towards designing labels in immersive environments to support situated visual search.

\subsection{Labeling Objects in AR}
In AR settings, we distinguish between labels for static and labels for dynamic objects that change over time.

Earlier studies have focused on managing label placement for static 3D landscape models~\cite{bellViewManagementVirtual2001} viewed from a top-down viewing angle.   
Subsequent work improved algorithmic performance to create embedded~\cite{maassDynamicAnnotationInteractive2006} and external labels~\cite{maassEfficientViewManagement2006} for landscape models in interactive 3D virtual space. 
Azuma et al.~\cite{azumaEvaluatingLabelPlacement2003} proposed a cluster-based algorithm to place dense occlusion-free AR labels of static in-view objects in real-time. 
Tatzgern et al.~\cite{tatzgernHedgehogLabelingView2014} designed a depth-encoded hedgehog labeling technique to connect labels to their targets. 
Using similar techniques, Madsen et al.~\cite{madsenTemporalCoherenceStrategies2016} studied the effect of label rendering space and update frequency on locating labels. Both studies used around ten labels for in-view objects on an AR tablet.

For dynamically changing scenes, Makita et al.~\cite{makitaViewManagementAnnotations2009} studied view management for annotating a small number of moving objects (less than five) in wearable AR devices. 
Orlosky et al.~\cite{orloskyHaloContentContextaware2015} resolved annotations overlapping with objects in AR using image detection and halo layout design for up to six labels. 
Grasset et al.~\cite{grassetImagedrivenViewManagement2012} proposed image analysis to place location-based external labels on mobile AR browsers for up to 30 labels, but working best with 10 labels to avoid visual clutter. 
Furthermore, Tatzgern et al.~\cite{tatzgernAdaptiveInformationDensity2016} proposed dynamic data clustering and filtering to manage a larger number of annotations with adaptive label displays in AR. 




\subsection{Out-of-View Object Visualization}
While labeling for out-of-view objects has only been preliminary explored, approaches for visualizing out-of-view objects have been widely studied.
A main difference between the two is that labels generally support a two-way retrieval workflow (i.e., object-to-label and label-to-object look-up), while out-of-view object visualization focuses on a one-way retrieval to allow users to find the object quickly.
Out-of-view object visualization can be classified into \textit{overview+detail}, \textit{focus+context}, and \textit{detail-in-context} techniques~\cite{cockburnReviewOverviewDetail2008}. 
For AR HMD environments \textit{detail-in-context} approaches are most suitable, which includes object labels.
%
\textit{Overview+detail} with an extra miniature map can introduce visual overload on the AR screen and \textit{focus+context} methods with view distortions~\cite{sarkarGraphicalFisheyeViews1992} can cause misinterpretation and difficulty in locating the target~\cite{zanellaEffectsViewingCues2002}.

\noindent \textbf{2D visualization.}
Previous \textit{detail-in-context} visualization approaches have used various visual proxies to encode spatial information for out-of-view objects for large graph navigation and notifications ~\cite{zellwegerCityLightsContextual,baudischHaloTechniqueVisualizing2003}.  
Wedge~\cite{gustafsonWedgeClutterfreeVisualization2008}, Halo~\cite{baudischHaloTechniqueVisualizing2003,burigatVisualizingLocationsOffScreen}, Arrow~\cite{burigatVisualizingLocationsOffScreen} and EdgeRadar~\cite{gustafsonComparingVisualizationsTracking2007} are well-known 2D visualization techniques that use abstract shapes as proxies to encode spatial information of off-screen targets.
A comparable concept for designing labels for out-of-view objects is encoding targets as insets on a map or large network visualization~\cite{carpendaleFrameworkUnifyingPresentation, ghaniDynamicInsetsContextAware2011, lekschasPatternDrivenNavigation2D}.
Ghani et al.~\cite{ghaniDynamicInsetsContextAware2011} designed dynamic insets to encode out-of-view targets on the boundary. 

\noindent \textbf{AR/VR visualization.}
Out-of-view object visualization techniques in AR/VR environments build upon 2D visualization techniques, such as 3D Arrows~\cite{chittaro3DLocationpointingNavigation2004,schinke2010visualization,tonnis2006effective,burigat2007navigation}, 3D Halo~\cite{trapp2011strategies}, attention funnel~\cite{bioccaAttentionFunnelOmnidirectional2006}, and radar projection~\cite{gruenefeldEyeSee360DesigningVisualization2017,jo2011aroundplot}.

Several user studies have applied and compared these techniques in target searching tasks in AR/VR.
Petford et al.~\cite{petfordComparisonNotificationTechniques2019} compare wedge visualizations to other attention-guiding techniques such as WIM~\cite{stoakleyVirtualRealityWIM1995} in a projector-based display environment for a single out-of-view target. Gruenefeld et al.~\cite{gruenefeldHaloWedgeVisualizing2018} further apply Halo~\cite{baudischHaloTechniqueVisualizing2003} and Wedge~\cite{gustafsonWedgeClutterfreeVisualization2008} techniques to AR and VR HMD and evaluate their performance in multiple (up to eight) out-of-view target search and direction estimation tasks.
Bork et al.~\cite{borkEfficientVisualGuidance2018} investigate visual guiding techniques for searching spatial virtual objects in MR with AR HMD. They compared six out-of-view object visualization techniques in a controlled user study, including four previous techniques, 3D Arrows~\cite{schinkeVisualizationOffscreenObjects2010}, AroundPlot~\cite{jo2011aroundplot}, EyeSee360~\cite{gruenefeldEyeSee360DesigningVisualization2017}, sidebARs~\cite{siuSidebARsImprovingAwareness2013}, and two novel techniques, 3D Radar and Mirror Ball, where users are asked to use the visualization to collect up to eight out-of-view objects that are spatially distributed. They found that 3D Radar and EyeSee360 are fastest when guiding users to collect objects outside FoV.
Gruenefeld et al.~\cite{gruenefeldComparingTechniquesVisualizing2019} used radar-like visualizations to encode out-of-view objects on AR HMDs. They compared different methods for showing locations for in-view objects, out-of-view objects, or both, but focused on the actual objects and did not use labels. They evaluated direction guiding and object selection tasks for eight objects.
Most existing out-of-view object visualization techniques suffer from occlusion and edge clutter with multiple out-of-view objects, and are not suitable to encode labels. They also primarily focus on attention guiding and searching for a single or a few out-of-view targets. Other visual analysis tasks such as value comparison have not been considered.

In our study, we explore both in-view and out-of-view object visualization with AR labels to support situated analytic tasks. Two in-view object labeling techniques  (\lsituated{} and \lboundary{}) are based on external labeling techniques. We designed three out-of-view object labels building upon previous out-of-view object visualization techniques. \lheight{} label extends the concept of dynamic insets~\cite{ghaniDynamicInsetsContextAware2011} and Wedge~\cite{gustafsonWedgeClutterfreeVisualization2008} to map an object's physical position onto the AR screen boundary. \langle{} label can be seen as a top-down view of radar projection mapped onto the AR screen boundary. \lvalue{} label encodes an object's data value and ranks labels by values to support situated analysis. 




\begin{figure*}[t!]
    \centering
    \includegraphics[width=\linewidth]{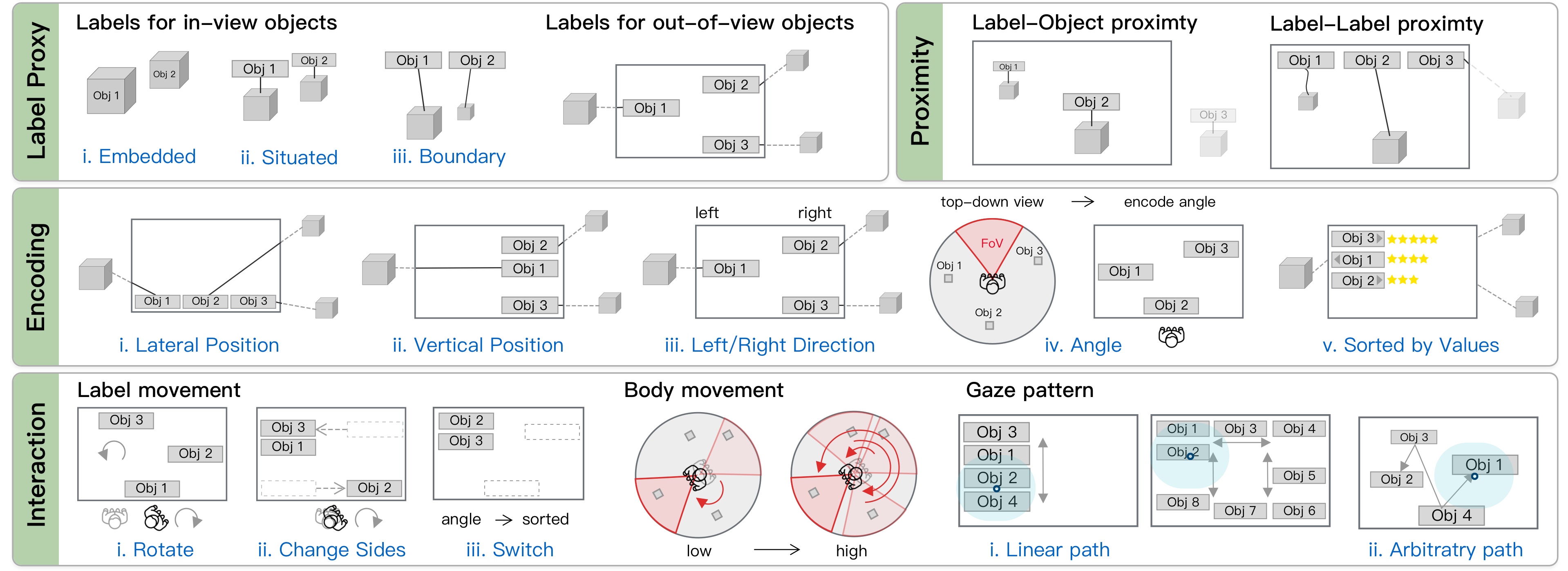}
    \caption{Label design space in immersive analytics. We categorize four factors of label design for situated target search, including label proxy, proximity, encoding, and interaction.
    } 
    \label{fig:designspace}
\end{figure*}

\section{Label Design Space \& Tasks}
\label{sec:design-space}

A \textit{label} is a visual representation of a text-based annotation attached to an \textit{anchor} on the target object. 
In the following, we discuss the most important properties of label design for immersive analytics, define goals and tasks for labels in situated search tasks, and outline some usage scenarios of AR labels. 


\subsection{Properties of Labels for Situated Analysis}
We define four distinct factors of label design in AR for situated analytics based on the relationship between labels, objects and the user, shown in Fig.~\ref{fig:designspace}. In addition, we identify other considerations that should be taken into account when designing labels in AR, such as object attributes and subjective design considerations.

Previous work~\cite{grassetImagedrivenViewManagement2012,steinDynamicLabelPlacement2008} has outlined rule-based properties of label layout in AR/VR scenes, such as that labels should be occlusion-free and placed close to the referred objects. While these are properties required to create a visually aesthetic and useful label layout, we focus on label properties that need to be considered in situated analytics tasks.

\textbf{Label Proxy.} Labels for objects in space need a visual representation. Proxies can include textual or visual information of the linked objects. Labels can be shown for visible objects (in-view), invisible objects (out-of-view), or both.
In immersive environments, a label can be \textit{embedded} inside the target or linked to the target through a leader line as an \textit{external} label. 
However, embedding labels accurately into objects can be computationally expensive due to the heterogeneous visual appearance of real-world objects and fluctuating view conditions~\cite{coelhoOsgarSceneGraph2004}. Furthermore, label embedding cannot be applied to out-of-view objects.
In our study, we focus on \textit{external} labels to explore a generalizable label design for real-world situated visual search tasks.

\textbf{Proximity.} Proximity describes the closeness between two labels or between a label and its object. Traditional AR/VR labels for in-view objects are optimized for label-object proximity. A higher proximity between label and object allows easier linking between the two. However, labels can be designed to prioritize label-label proximity to support efficient visual comparison, such as comparing the values on all labels. 

\textbf{Encoding.} Label positions can encode additional attributes such as the lateral or vertical position of an object with respect to the user's orientation/gaze. Alternatively, a label position can also encode the left/right direction (i.e., whether the object is to the left or to the right of the current view), angle (i.e., the angle between the user's view direction and the object), or the data value of the object.

\textbf{Interaction.} To support spatial search, labels are designed to respond to a user's interaction and movement. 
Label movement describes if the label placement remains static or updates based on user movement. On the other hand, to use the labels in visual search tasks, the user has to perform body or gaze movements. Body movement reflects the degree of the user's physical movement required in each label design; gaze pattern classifies the direction of gaze movement in order to access the labels, including linear and arbitrary paths. For example, a user's gaze pattern is linear if labels are aligned vertically or horizontally, and arbitrary if labels are situated at the object's position.

\textbf{Other Considerations for AR label design.} 
In addition to the factors outlined above, there are other considerations that influence the design of efficient labels that focus on \emph{object attributes} and \emph{subjective measures}. 
\textbf{Object attributes} include \emph{the number of objects} (sparse versus dense), \emph{the spatial distribution of objects} (full 3D space versus restricted to certain areas such as the half-space above ground), and \emph{object movement} (static versus dynamic).
%
\textbf{Subjective measures} include the user's \emph{familiarity} with a chosen label design as well as the \emph{predictability} of label locations. For example, if labels are always located at the lower screen boundary, users will have an easier time locating them.



\subsection{Goals \& Tasks}
\label{sec:goal_task}
Inspired by previous work in AR information retrieval and situated analytics~\cite{elsayed_situated_2015,bioccaAttentionFunnelOmnidirectional2006,buschelHereNowRealityBased2018,bachDrawingARCANVASDesigning}, we have identified two primary goals of label design in AR and for situated visual search tasks.

\noindent
\textbf{G1 - visualize the spatial relationship between the objects and the user to support situated spatial search.} Users should be able to explore all objects and locate a target in space more easily with the provided labels, for in-view as well as out-of-view objects.

\noindent
\textbf{G2 - provide extra information for objects to support visual analysis tasks in the real world} that would otherwise be extremely tedious or impossible to perform. Such visual search tasks include but are not limited to identifying outliers, comparing multiple targets, or making overall estimations.

We identify common visualization tasks to support situated spatial and visual search. 
Based on the typology of abstract visualization tasks by Brehmer and Munzner~\cite{brehmerMultiLevelTypologyAbstract2013} \textit{``the user must find elements of interests in the visualization"}. 
Labels in AR already encode an object's \textit{identity} and point towards the \textit{location} of the referred object. 
Therefore, labels can be seen as the result of an information retrieval process~\cite{yoo3DUserInterface2010,buschelHereNowRealityBased2018}, and situated visual search requires interpreting available labels in the AR scene.
According to Brehmer and Munzner's typology~\cite{brehmerMultiLevelTypologyAbstract2013},
labels for situated analysis need to support the three low-level querying actions, \tidentify{}, \tcompare{}, and \tsummarize{}. 

\noindent
\textbf{T1: \tidentify{} a single target among all objects of interest.} Users need to be able to identify a single target from all objects of interest in the environment. Typically, users need to physically examine each object in the environment to decide on a single target. Labels in AR can provide a quick overview of all objects to help identify a single target (e.g., a product with the lowest price), and spatial cues to guide the user to the desired target.

\noindent
\textbf{T2: \tcompare{} between multiple targets.} Users need to perform comparisons between multiple targets in the environment. Typically, users have to examine and memorize the object-specific data for each object. With labels, this task can be done with visual comparison from the labels, such as comparing the prices and calories of similar products. 

\noindent
\textbf{T3: \tsummarize{} data across all targets.} To get an overview of the scene or perform comparisons at a larger scale, users need to summarize data across targets. This task is very challenging in a real-world scenario, such as comparing the average prices of one brand against another in a store, or house prices in two neighborhoods.
With labels, users can summarize the trend or identify outliers among all available targets.

\subsection{Usage Scenario}
\label{sec:usage_scenario}
We exemplify the identified tasks of Sec.~\ref{sec:goal_task} with real-world usage scenarios of AR labels. 

\textbf{Grocery shopping.} A person shopping in the supermarket needs to know the location of each item on the shopping list. Instead of looking for items aisle by aisle, they look up the labels on an AR HMD. Labels show the categories of items and encode the directions of each category that is outside of the current view. 
They \textit{identify} ``Fruits" and follow the label to directly locate the aisle. Similar to Fig.~\ref{fig:teaser} (c), AR labels show the price of different fruits and varieties with color-coded bar charts. The shopper \textit{compares} their prices right in place without having to examine each item. They also estimate the price difference between apples and tomatoes by \textit{summarizing} the prices shown on the labels. With the labels to support situated analysis, they can make data-driven decisions without much effort.

\textbf{Spatial navigation.} A traveler standing in front of a tourist site is deciding where to go for lunch in the surrounding area. The search results are shown as labels on the AR HMD and are encoded with directions. Among all labels, the tourist \textit{summarizes} the average ratings of Japanese and Italian restaurants. They decide on Japanese food and use the labels to \textit{compare} the prices of individual restaurants. While making a decision, they are able to continue to explore the surrounding area. Finally, they \textit{identify} the target restaurant and follow the label direction to the destination.

\section{Label Placement Design}
\label{sec:label-placement-design}
Based on our label design space, we explore different combinations of label placement properties as shown in Table~\ref{table:1}, including two variations of labels for in-view objects -- \lsituated{} and \lboundary{}, and three variations of labels for out-of-view objects -- \lheight{}, \langle{} and \lvalue{}. 
Each label design consideration is discussed below and shown in Fig.~\ref{fig:design-conditions} and the supplemental video.

\subsection{\lsituated{}}
\label{sec:situated}
 \lsituated{} is the most common labeling technique in existing AR/VR applications~\cite{bellViewManagementVirtual2001, maassEfficientViewManagement2006, grassetImagedrivenViewManagement2012},  such as a name tag and health bar placed above an avatar in a video game. We design \lsituated{} labels as a baseline condition in our user study.
\lsituated{} labels are placed directly above their referred objects with a straight leader line connecting the object center to the bottom center of the label (Fig.~\ref{fig:design-conditions} (1)). 
When objects are too close, we adjust the vertical label positions to avoid overlaps.
The close proximity between labels and objects makes it clear to which object a label is connected to and reduces time and effort to identify the object. 
Due to the familiarity of similar \lsituated{} labels in existing 3D environments, it feels natural to users. 
Therefore, \lsituated{} labels have the advantage of \textbf{high label-object proximity} and \textbf{high familiarity}. 

On the other hand, \lsituated{} labels require \textbf{high body movement} because the label placement does not encode any spatial information of the objects. It requires users to do a full-space scan to search for out-of-view objects. When scanning the entire surrounding space, it is also difficult to find labels because they appear at different vertical heights and require more \textbf{arbitrary gaze patterns}. In particular, when comparing multiple spatially sparse objects, users have to remember the information on the labels as well as the locations of the corresponding objects. This \textbf{low predictability} of label locations makes the comparison between multiple targets error-prone and inefficient.




\subsection{\lboundary{}}
\begin{figure*}[t!]
    \centering
    \includegraphics[width=\linewidth]{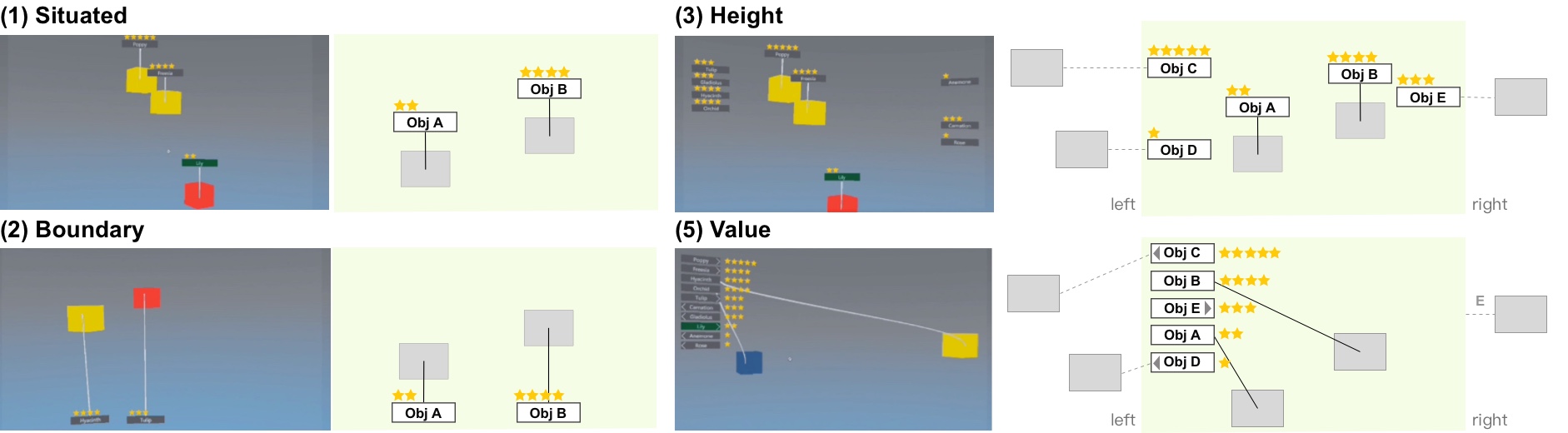}
    \caption{AR screen views (left) and diagrams (right) of the four label design in Sec.~\ref{sec:label-placement-design}. \langle{} design is shown in Fig.~\ref{fig:teaser} (a) and (b).
    } 
    \label{fig:design-conditions}
\end{figure*}

\textsc{Boundary} labels are an alternative design of in-view labels, with a focus on visual search and comparison tasks. As shown in Fig.~\ref{fig:design-conditions} (2), we align 
\lboundary{} labels horizontally at the bottom edge of the AR screen and encode the horizontal positions of the linked in-view objects. We prioritize objects at the center of the view and push overlapping labels towards the left or right to avoid overlaps when necessary.
The one-dimensional label placement makes it easy to quickly scan all labels in one direction without having to change the vertical gaze pattern to find labels. 
Labels are also closer together, which makes label-to-label comparison more efficient. This helps users to get a faster overview and reduces gaze movement. Label locations are more predictable as labels are always placed at the bottom. 

We originally designed the labels to be aligned at the top, but initial user feedback suggested that users found it uncomfortable to constantly have to look up with their eyes. Therefore, we place boundary labels at the bottom to mimic the conventional task bar or dock placement on a computer or mobile device.
Summarizing, \lboundary{} labels have the advantage of \textbf{label-label proximity}, a \textbf{linear gaze pattern}, and \textbf{high predictability}.

Similar to \lsituated{}, \lboundary{} labels also require \textbf{high body movement} to complete visual search and comparison tasks. Furthermore, they require a separate step to link the label to the object due to \textbf{lack of label-object proximity}. The longer leader lines make it more difficult for users to retrieve the linked object and can cause visual clutter when there are multiple objects in view and lines are overlapping with the objects in the scene.  





\begin{table}[t!]
    \centering
        \begin{tabular*}{\columnwidth}{ p{1em}| c c c c c} 
         Factor & \textit{Situated} & \textit{Boundary} & \textit{Height} & \textit{Angle} & \textit{Value} \\ 
         \hline\hline
         \textbf{Label} &&&&& \\
         \multicolumn{1}{r|}{In-View} & \cmark & \cmark & \cmark & \cmark & \cmark \\ 
         \multicolumn{1}{r|}{Out-of-View} & \xmark & \xmark & \cmark & \cmark & \cmark \\
         \hline
         \textbf{Proximity} &&&&& \\
          \multicolumn{1}{r|}{Label-Label} & low  & high & low/med.  & high & high \\
          \multicolumn{1}{r|}{Label-Object}  & high & med. & high & med. & low \\
          \hline
         \textbf{Encoding} &&&&& \\
          \multicolumn{1}{r|}{Lateral position}  & \cmark / - & \cmark / -  & \cmark / \xmark & \cmark & \xmark \\
          \multicolumn{1}{r|}{Vertical position}  & \cmark / - & \xmark / - & \cmark & \xmark & \xmark \\
          \multicolumn{1}{r|}{Direction}  & \xmark & \xmark & \cmark & \cmark & \cmark \\
           \multicolumn{1}{r|}{Angle}  & \xmark & \xmark & \xmark & \cmark & \xmark \\
          \multicolumn{1}{r|}{Value}  & \xmark & \xmark & \xmark & \xmark & \cmark \\
           \hline
          \textbf{Interaction} &&&&& \\
          \multicolumn{1}{r|}{Label movement}  & \xmark & \xmark & \cmark & \cmark & \xmark \\
           \multicolumn{1}{r|}{Body movement}  & high & high & med. & low & med. \\
           \multicolumn{1}{r|}{Gaze pattern}  & arb. & linear & arb./lin. & linear & linear \\
           \hline \hline
           \textbf{Subjective} &&&&& \\
           \multicolumn{1}{r|}{Familiarity} & high & low & med. & low & med. \\
           \multicolumn{1}{r|}{Predictability} & low & high & low & high & high \\
         \hline
         & \multicolumn{5}{l}{*arb. = arbitrary, lin. = linear, med. = medium}
        \end{tabular*}
    \caption{ Characteristics (rows) of five AR label placement designs (columns) in our study. The slash indicates different properties of labels for in-view and out-of-view objects.}
    \label{table:1}
\end{table}

\subsection{\lheight{}}
In our study, we use three encodings for out-of-view object labels. Compared to label design for in-view object labels, they all require less body movement and support a linear gaze pattern to track labels, but more labels also result in higher visual density.

As shown in Fig.~\ref{fig:design-conditions} (3), \lheight{} labels are placed at the left and right boundaries of the view, based on the closer lateral direction of the target object. Furthermore, their y position encodes the relative \emph{height} at which the linked object will come into the screen as the user rotates. 
Showing labels for all available objects in the scene allows users to search for and compare targets without physically moving around. The height-encoded label placement provides cues to both the lateral direction and vertical position of an object, and thus users can know immediately which way to turn to when searching for a single target. For example, seeing a label in the top left border means the user will turn left and look upward to find the target. When the object appears in view, the height-encoded label will become \lsituated{} and directly lead the users to the target object. 
\lheight{} labels provide an \textbf{encoding for the vertical position and lateral direction} of out-of-view objects and exhibit \textbf{high label-object proximity}. 

Since \lheight{} labels are arranged along the left and right borders of the view, based on which border the target object is closer to, their placement on the screen is changing as people turn. For example, objects that were closer to the left border might become closer to the right border as the user rotates. 
While \textbf{label movement} provides real-time directional cues on objects, it requires users to search for and constantly track out-of-view labels (i.e., labels for out of view objects) while rotating and moving their gaze. 
Due to the lack of distance information (i.e., how far outside the field of view an object is), users have to follow a label at all times to find the object, which can cause higher mental stress and slightly \textbf{more body movement}. Furthermore, when comparing multiple out-of-view objects, it takes extra mental effort to search for and compare labels, as they are not ordered by angles or values.



\subsection{\langle{}}
\langle{} labels are placed along all four edges of the user's view boundary and encode the \emph{angular} direction of all available objects as seen from a top-down view, as shown in Fig.~\ref{fig:teaser} (b). The top boundary represents the front-facing direction (i.e., the direction the user is currently looking at) and the bottom boundary represents the area behind the user.  
Therefore, all labels for in-view objects are naturally placed along the top boundary and linked to their target objects with leader lines. Labels for out-of-view objects are placed along the left, bottom, and right view boundary, depending on their position in space.
As users turn around, the \langle{} labels rotate along the boundary to reflect the relative angular direction of the objects to the user. Therefore, \langle{} provides a precise encoding of the angular distance between the user and the target object, as well as between objects. 
This can increase the spatial awareness of users and reduces mental and physical efforts when locating objects.
Similar to \lboundary{}, \langle{} label placement is very predictable and aligned vertically and horizontally, which allows easier label-to-label comparison. 
As users move, \langle{} labels rotate in either clockwise or counter-clockwise direction and hence do not distract users as much as in the \lheight{} condition.
\langle{} labels provide an encoding for an object's \textbf{lateral direction and angle}, and exhibit \textbf{label-label proximity}, and \textbf{high predictability}.

On the other hand, users might have \textbf{low familiarity} with angle-encoded labels, which can be difficult to interpret without prior training. When looking for a target object, the longer leader lines between label and object might cause more cognitive effort to link the object and the label due to \textbf{lack of label-object proximity}. 
Similar to \lheight{} labels, when comparing multiple labels, it takes time to find and sort the label values.




\subsection{\lvalue{}}
As shown in Fig.~\ref{fig:design-conditions} (5), \lvalue{} labels are placed stationary along the left view boundary and ordered vertically by the label's \emph{value}. In order to stack more labels and allow more direct value comparisons similar to a horizontal bar chart, we chose a parallel label layout where the label's value is shown to the right of the label's text.
We indicate the direction of the linked objects with left/right arrow icons on the labels. When the linked object is in view, the arrow icon disappears and a leader line appears.
Label placement is very predictable as people only have to look at the left boundary to look for labels and then follow the arrow to find the object. When comparing multiple labels, \lvalue{} has the advantage of efficient value comparisons due to the fixed label layout and implicit ranking. Users only have to move their gazes vertically to make cross-label comparisons.
Therefore, \lvalue{} labels provide an encoding for an object's \textbf{value and direction}, and exhibit \textbf{label-label proximity}, and \textbf{high predictability}.

While \lvalue{} labels have a stable and predictable placement, it requires an extra step to interpret the arrow icons and follow the leader line to find the objects due to \textbf{lack of label-object proximity}. Encoding spatial information in icons is more subtle than encoding it as label positions, which might require more mental effort to interpret and \textbf{more body movement} than \langle{} labels. Furthermore, leader lines of different labels might exhibit line crossings due to the label's vertical position being independent of the object's vertical position. 




\subsection{AR Interface Implementation}
We developed and implemented all label designs using Unity v2020.1.12~\cite{unity} and the Mixed Reality Toolkit v2.5.1~\cite{MRTK}. In order to vary object distribution and achieve higher tracking fidelity of the objects in the study, we use VR to simulate the real-world object space and AR screen. 
Simulating an AR environment with VR is a common method to create a controlled environment in a user study \cite{marquardtComparingNonVisualVisual2020,raganSimulationAugmentedReality2009,jungEnsuringSafetyAugmented2018}. 

We set up our AR interface as a semi-transparent canvas placed $1.8m$ in front of the user with a $35^{\circ}\times25^{\circ}$ FoV. We display the AR screen in the users' FoV at a fixed location, and, therefore, the labels on the AR screen fall on the periphery of the user's vision (see Fig.~\ref{fig:teaser} (a)). 
Our chosen AR screen size falls within the FoV of the current state-of-the-art AR HMDs, Magic Leap 1 ($40^{\circ}\times30^{\circ}$) and HoloLens ($43^{\circ}\times29^{\circ}$).
We apply the tooltip component of the Mixed Reality Toolkit UX building block~\cite{mrtk-tooltip}. The leader lines are designed to curve based on the depth between the object and the label on the AR screen to provide a sense of depth, such as in Fig.~\ref{fig:design-conditions} (5).
We place all objects around the user in 360 degrees. Details on data generation are discussed in Sec.~\ref{sec:data}. 

\section{User Study}
\label{sec:user-study}
We evaluate the usefulness of our five label designs in supporting situated visual search tasks in a controlled user study.
In particular, we want to investigate in which situations out-of-view labels are helpful, whether in-view labels have any advantages, and the trade-offs between the two.

\subsection{Experimental Design}

\noindent \textbf{Labeling conditions.}
We evaluated the five label conditions described in Sec.~\ref{sec:label-placement-design}, including \lsituated{}, \lboundary{}, \lheight{}, \langle{}, and \lvalue{}. In all conditions, each label shows the unique name of the object and a rating ranging from 1 to 5 stars. 

\noindent \textbf{Data sizes.} 
We generated datasets with two data sizes -- 10 and 20 objects. 
Previous studies on label design for in-view objects in tablet AR used around 10 labels~\cite{tatzgernHedgehogLabelingView2014,madsenTemporalCoherenceStrategies2016}. 
We also constrain the amount of information in the AR FoV to avoid information overload and occlusion~\cite{redelmeierAssociationCellularTelephoneCalls1997,pascoal2017information}.
10 and 20 objects are chosen for different real-world use cases, such as comparing a fixed set of targets and exploring a broader set of options, respectively. 


\noindent \textbf{Experiment Set-Up.}
We conducted our study in an indoor lab space of approximately $385 ft^2$.
We used an Oculus Quest virtual reality headset, with a $96^{\circ}\times94^{\circ}$ FoV, $1440\times1600$ pixel resolution, and a weight of $571g$ to render objects, labels, and the semi-transparent AR screen. 
The headset was connected to a PC through a $5m$ USB3 Type-C cable. The PC had an Intel i7-9700F 3.00GHz processer and NVIDIA GeForce RTX 2070 graphics card.  

To reduce motion sickness and maintain orientation in VR, we rendered a $4\times4$m virtual floor and an arrow at the center of the virtual floor indicating the user's starting position (i.e., front-facing direction). 
The study had a stationary set-up and required minimum body movement. Users did not need to walk around but only had to turn, move their head and use one controller to interact with the question pop-up windows. Participants were required to return to the front-facing direction before each trial.
 
\subsection{Task \& Data}
\label{sec:data}
We used three visual search tasks (see Sec.~\ref{sec:goal_task}) based on the typology of abstract visualization tasks by Brehmer and Munzner~\cite{brehmerMultiLevelTypologyAbstract2013}.

\noindent \textbf{\tidentify{}:} 
\textit{What is the color of the object linked to the green label?} 
We colored the target object's label in green and all others in grey. All objects were randomly colored in either red, yellow, or blue. Participants had to identify the color of the object linked to the green label and select the answer in the pop-up menu. 

\noindent \textbf{\tcompare{}:} \textit{What is the color of the object with the highest rating?}
Three labels were colored in green, all others in grey. 
All objects were randomly colored in either red, yellow or blue, while each of the three linked objects was assigned a different color. Participants had to compare the ratings of the three green labels and find the color of the object with the highest rating. 

\noindent \textbf{\tsummarize{}:} \textit{In the two colored clusters, which cluster has a higher average rating?} We designed two spatially separate clusters, with three objects in the red cluster and four objects in the blue cluster. Initially, the clusters were not shown, and all objects were colored in grey. Two selected labels were colored in red and blue, the others in grey. 
\begin{wrapfigure}{r}{0.18\textwidth}
\includegraphics[width=\linewidth]{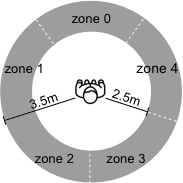} 
\caption{Object locations.}
\label{fig:zone}
\end{wrapfigure}
Participants first had to locate and click on the two objects linked to the blue and red labels to reveal the clusters. 
The labels and the objects of each cluster were colored in red and blue. They then determined which cluster had the higher average rating: red, blue or equal.
The task represents a potential user flow of identifying two targets of interest and expanding the search to their immediate neighborhood.

\begin{figure}[t!]
    \centering
    \includegraphics[width=\linewidth]{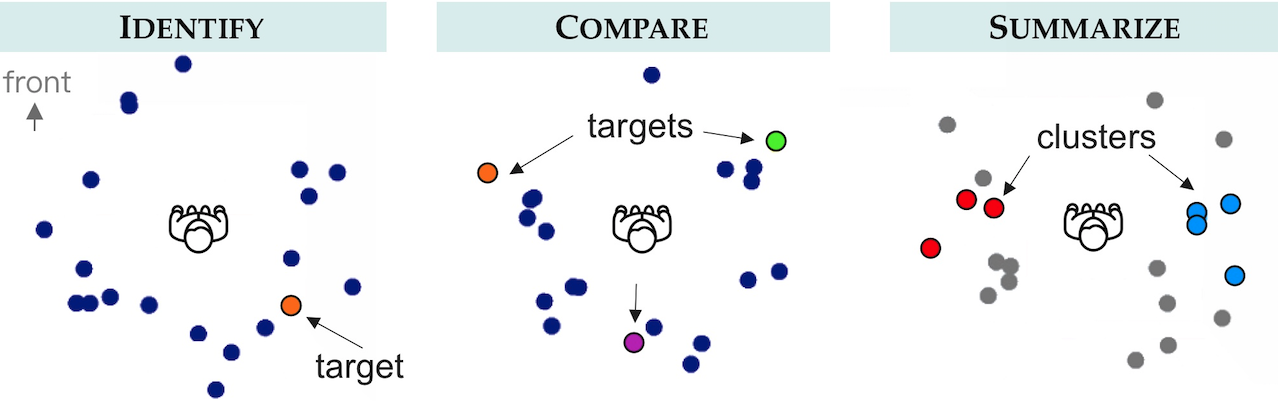}
    \caption{Example datasets for all three user tasks showing the spatial distribution of objects. Each small circle represent the object's x and y location in the trial in a top-down view (object height is not shown). The selected target(s) for each task are highlighted.} 
    \label{fig:dataset}
\end{figure}

We generated a distinct dataset with 10 or 20 spatially sparse objects for each trial in our study.
Each object is represented as a $20cm\times20cm$ cube, with a minimum distance of $40cm$ between cubes. All objects were randomly placed within $2.5m$ to $3.5m$ from the user at a viewing height between $0.5m$ to $2m$.
To distribute the objects in space, we divided the circular space around a user into five zones (see Fig.~\ref{fig:zone}) and ensured that each zone contained a similar number of objects.
Each object's label was randomly assigned a unique name selected from a list of items, such as fruits and flowers, and a rating value between 1 and 5. To control for the different reading speeds of our participants, we used color to highlight the targets in our tasks instead of relying on text labels.

We show sample data for each task in Fig~\ref{fig:dataset}.
For the \tidentify{} task, we randomly selected an object in zone 2 or 3 as the target object to avoid showing the target immediately in front of the user. 
For the \tcompare{} task, we randomly selected three objects from different zones and assigned a non-repetitive rating value from 1 to 5 and a distinct color from red, yellow and blue. We intentionally spread out the target objects in space so that no two objects being compared could be seen in the AR view at the same time.
We also did not assign 5-star ratings to the targets being compared during the trials to make the task more challenging.
For the \tsummarize{} task, we selected two objects from different zones to be labeled with red and blue labels at the beginning of the task. We selected objects close to the red and blue object as the red and blue clusters respectively, which resulted in three red objects and four blue objects in each trial. We then assigned rating values to each object so that the resulting average of each cluster was an integer. 

\subsection{Participants \& Procedures}
\noindent \textbf{Participants.} We recruited 15 participants from the University mailing list. Due to the COVID-19 pandemic and restrictions to campus access, we limited participation to university students and staff only. Participants ranged in age from 18-34 years. Five identified as female and ten as male. Seven participants had prior experience with VR HMDs, and one had experienced AR HMDs.

\noindent \textbf{Procedure.}
We followed a full-factorial within-subject study design to account for individual differences between participants. We balanced the order of label conditions using a Latin square (5 groups), and fixed the task order by increasing task complexity: \tidentify{}, \tcompare{}, and \tsummarize{}.

Each condition (label $\times$ task) included 2 training trials and 6 timed study trials. Each participant completed 90 study trials: 5 label conditions $\times$ 6 trials $\times$ 3 tasks.
Participants first filled out a consent form and were introduced to the study procedure and all five label designs with slides. 
Next, the instructor helped participants set up the VR headset and showed them how to select an answer with the trigger button. 
Before each task, the instructor introduced the task and reminded participants to perform the task \textit{as precisely and as quickly} as possible during the timed trials.
Participants were encouraged to spend as much time as needed on training. 
They could take breaks between each trial and task.
After each task, the instructor collected participants' oral feedback. After the final task, participants filled out a post-study questionnaire. 
The whole study took 1.5 hours to complete, and each participant was compensated with a $\$20$ gift card.

\subsection{Measures}
We recorded the \textit{time} performance of each trial from starting the visualization to the time participants double-clicked to input their answers. 
For \textit{accuracy}, we compared their answer to the correct answer. 
We also collected participants' subjective feedback for each task and their overall evaluation in the end with a post-study questionnaire, including a standard NASA-TLX survey, qualitative feedback, and subject rankings for the five labeling conditions.
To extract insights from the qualitative feedback, we derived a set of codes in an open coding session among three authors for the first five participant responses.
The first author then applied the set of codes to the feedback of the remaining 10 participants.

\subsection{Statistical Analysis}
For dependent variables or their transformed values that can meet the normality assumption (i.e., time), we used \emph{linear mixed modeling} to evaluate the effect of independent variables on the dependent variables~\cite{Bates2015}. 
Compared to repeated measure ANOVA, linear mixed modeling is capable of modeling more than two levels of independent variables and does not have the constraint of sphericity~\cite[Ch.\ 13]{field2012discovering}.
We modeled all independent variables (i.e., label placement techniques and data sizes) and their interactions as fixed effects. A within-subject design with random intercepts was used for all models. 
We evaluated the significance of the inclusion of an independent variable or interaction terms using log-likelihood ratio. 
We then performed Tukey's HSD post-hoc tests for pair-wise comparisons using the least square means~\cite{Lenth2016}. 
We used predicted vs. residual and Q~---~Q plots to graphically evaluate the homoscedasticity and normality of the Pearson residuals, respectively. 
For dependent variables that cannot meet the normality assumption (i.e., accuracy, NASA-TLX ratings), 
we used a \emph{Friedman} test to evaluate the effect of the independent variable, as well as a Wilcoxon-Nemenyi-McDonald-Thompson test for pair-wise comparisons. Significance values are reported for $p < .05 (*)$, $p < .01 (**)$, and $p < .001 (***)$, respectively, abbreviated by the number of stars in parenthesis.



\section{Results}
\label{sec:results}
\begin{figure}[t!]
	\centering
	\includegraphics[width=0.95\columnwidth]{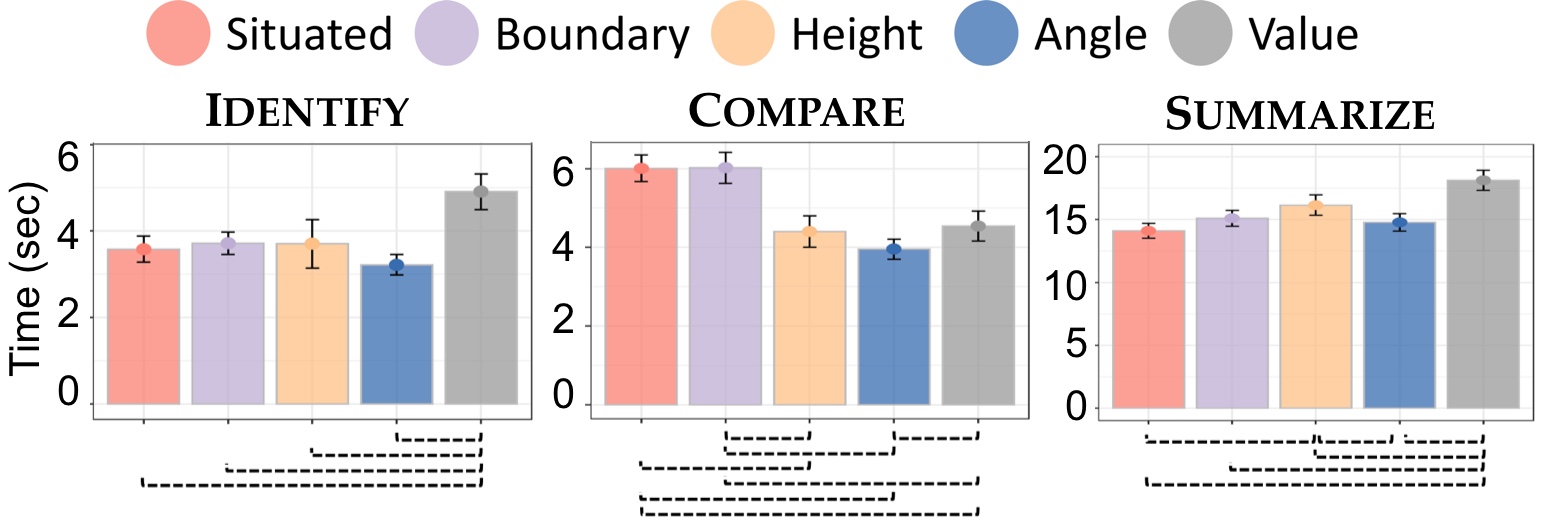}
	\caption{Results of task completion \emph{time} by tasks. Confidence intervals indicate 95\% confidence for mean values. Dashed lines in all charts indicate statistical significance for $p<.05$. 
	All out-of-view labels outperformed in-view labels in the \tcompare{} task.
    }
	\label{fig:study-results}
\end{figure}

\begin{figure}[t!]
	\centering 
	\includegraphics[width=0.9\columnwidth]{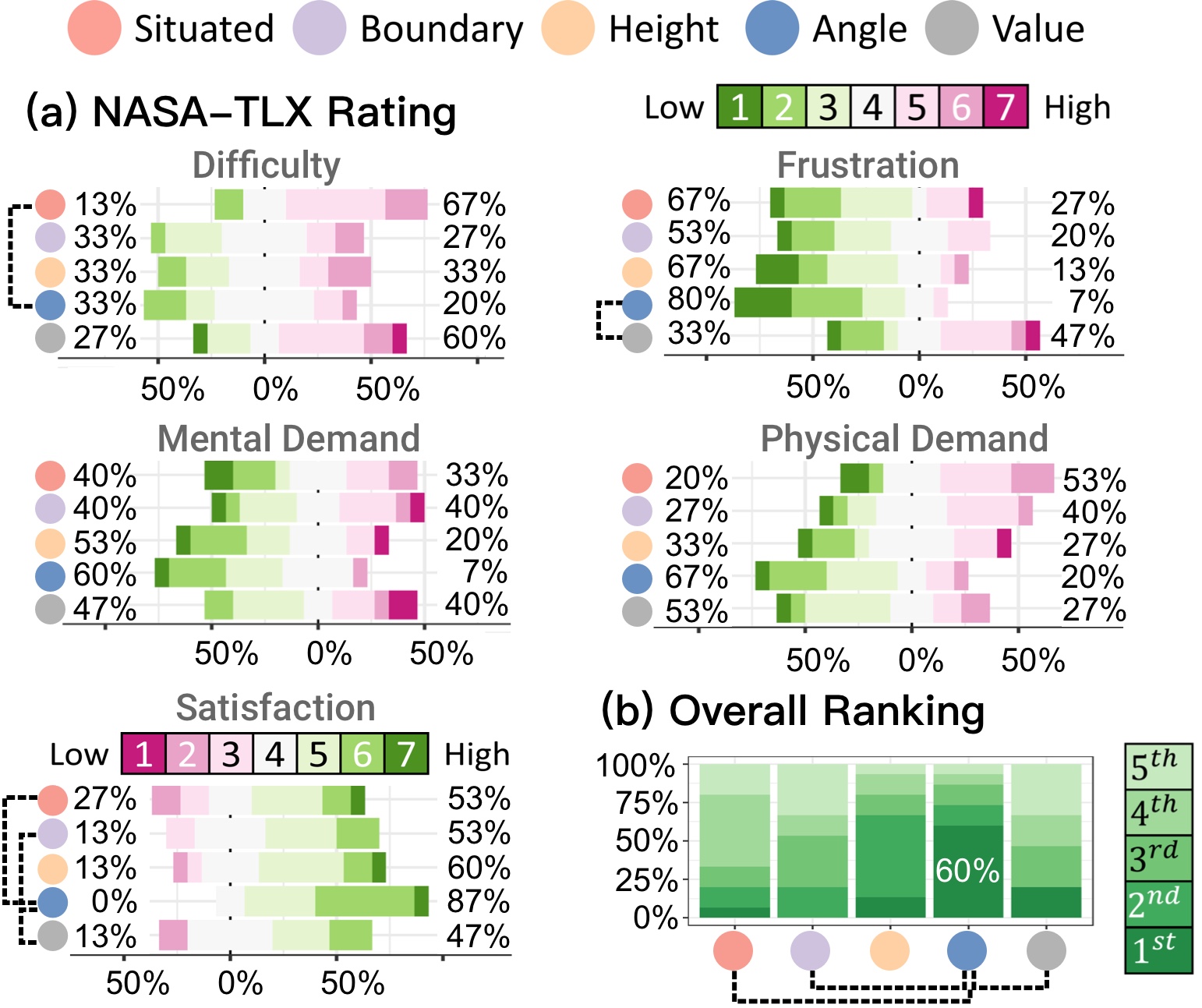}
	\caption{(a) \emph{NASA-TLX ratings}. The percentage of negative (pink) and positive ratings (green) is shown next to the bars. (b) User preference ranking for each labeling condition. All dashed lines indicate $p<.05$.}
	\label{fig:study-result-ranking}
\end{figure}

We did not find significant effects of label placement techniques and data sizes on accuracy. User accuracy was similar and high across all conditions (see supplemental material).
Therefore, we focus our analysis on time (see~\autoref{fig:study-results}), subjective ratings, rankings (see~\autoref{fig:study-result-ranking}), and qualitative feedback.

\subsection{Completion Time}
\noindent\textbf{\tidentify{}}:
We found that label placement techniques had a significant effect on task completion time in \tidentify{} ($***$).
\lvalue{} was slower than all other techniques (all $***$).
There was no difference between label techniques for in-view and out-of-view objects. 
We found no significant effect in data size and the interaction between the two factors (i.e., label placement technique $\times$ data size) in \tidentify{}.

\noindent\textbf{\tcompare{}}:
We also found that label placement techniques had a significant effect on task completion time in \tcompare{} ($***$).
Having labels for out-of-view objects (\lheight{}, \langle{}, and \lvalue{}) was faster than techniques that only provided labels for in-view objects (\lsituated{} and \lbound{}), all $***$.
For labels of out-of-view objects, \langle{} was faster than \lvalue{} ($*$), and \langle{} tended to be faster than \lheight{}, but not statistically significant ($p=0.060$).
For labels of in-view objects, there was no difference between \lsituated{} and \lbound{}.
Data sizes had a significant effect on time in the \tcompare{} task ($***$) with large data sizes increasing completion time. 
No significant effect was found in the interaction between the two factors.

\noindent\textbf{\tsummarize{}}:
Label placement techniques had a significant effect on task completion time in \tsummarize{} ($***$).
Like in \tidentify{}, \lvalue{} was slower than other techniques (all $***$).
We also found \lheight{} was slower than \lsituated{} ($***$) and \langle{} ($*$).
No significant effect was found in data size, and the interaction between the two factors was marginally significant ($p=0.057$).


\subsection{Subjective Ratings and Ranking}
Our NASA-TLX results (\autoref{fig:study-result-ranking} (a)) show that label placement techniques had a significant effect on the perceived \emph{satisfaction} ($**$), \emph{difficulty} ($*$), and \emph{frustration} ($*$).
\textbf{\langle{} received higher satisfaction ratings than \lsituated{}, \lboundary{}, and \lvalue{} ($**$)}. \lsituated{} was rated more difficult than \langle{} ($*$). \lvalue{} was rated more frustrating than \langle{} ($*$).
We also found that placement techniques had a significant effect on a user's overall ranking (\autoref{fig:study-result-ranking} (b)). \langle{} was more preferred than \lsituated{}, \lbound{}, and \lvalue{} (all $*$).

\subsection{User Strategies for each Task}
We observed substantial differences in user strategies 
for different label conditions apart from the task performance. 

\noindent \textbf{\tidentify{}.} 
Participants used different strategies for in-view and out-of-view object labeling conditions.
In \lsituated{} and \lboundary{}, all 15 participants chose an arbitrary direction to start a full-space search to find the target object. 
In \lheight{}, \langle{}, and \lvalue{}, all participants scanned the AR screen for the target label before following the directional cues on the label to find the target. 


\noindent \textbf{\tcompare{}.} 
In \lsituated{} and \lboundary{}, participants had to physically locate all target objects and memorize the label values when moving around. 
There was also a slight difference between them, where 
In \lsituated{}, participants had to constantly move their gaze up and down as they turned to find labels.
In \lboundary{}, participants only moved their gaze horizontally to scan for labels and vertically to link to the object.
With labels for out-of-view objects,
users first compared labels and directly decided on the final target without having to physically turn around. 



\noindent \textbf{\tsummarize{}.}
In label conditions for in-view objects, all participants sequentially found clusters and calculated their group average one by one, which required them to remember the average of the previous cluster. Sometimes participants had to return to the first cluster to double-check their answer.
In out-of-view object labeling conditions, users directly looked at the labels to check the values of both clusters and estimate cluster averages. 
However, for the \lheight{} condition, four participants reported that finding labels on the screen took more work than physically turning back to the other cluster because label positions were less predictable and not grouped into clusters. They performed the task exactly the same as with \lsituated{}.

\subsection{Overall Qualitative Feedback}

Participants commented on the overall usefulness of each label condition for visual search in the post-study questionnaire.

\noindent
\textbf{Labels for In-View Objects.}
With \lsituated{}, 9 out of 15 participants commented that it was easy to link label to object. 3 of them also appreciated the lower visual density. 
With \lboundary{}, participants cited various advantages, including predictable label locations (4), easy matching from label to in-view objects (4), being good for comparison (3), label-to-label proximity (2), and being good for summarize tasks (2). 
\textbf{The major downside of labels for in-view objects was that they require a full spatial scan}, 
which was reported by 13 (\lsituated{}) and 7 (\lboundary{}) participants.
Participants found that it was physically demanding (6 and 3), had a high load on memory (5 and 3), and was bad for compare tasks (4 and 2) and summarize tasks (3 and 2). 
They also mentioned that is was difficult to find objects for \lsituated{} (2) and leader line clutter for \lboundary{} (5).

\noindent
\textbf{Labels for Out-of-View Objects.} 
Getting an immediate overview of objects and ratings was reported as an advantage in all three out-of-view object labeling conditions, with 5, 4, and 6 participants for \lheight{}, \langle{}, and \lvalue{}, respectively. 
\textbf{Good directional cues was the major advantage for \lheight{} and \langle{} labels}, which was reported by 7 and 9 participants, respectively. In particular, 5 out of those 9 appreciated the precise directional cues provided by \langle{} labels. Participants also found it is easy to go from label to object in both conditions, 3 in \lheight{} and 6 in \langle{}. 
For the \lheight{} condition, participants reported predictable height locations of objects (2) and being good for comparisons (2). \langle{} was also preferred for its predictable label locations (1), similar to \lboundary{}.
\textbf{\lvalue{} was found most beneficial for comparison tasks.} Participants found  \lvalue{} good for comparisons (6), and good for compare tasks (3) and summarize tasks (6), and for providing good directional cues (2).

\noindent
\textbf{Leader line clutter was a common disadvantage reported for \lboundary{}, \langle{}, and \lvalue{}.} The three conditions with longer leader line were reported to have leader line clutter, 5 in \lboundary{} and \langle{}, and 7 in \lvalue{}. Participants also reported that it took time in the \langle{} condition to learn how to interpret the label placement (5).
Unlike \langle{}, which showed the precise direction of objects, participants reported that \textbf{disadvantages for the \lheight{} and \lvalue{} conditions were subtle directional cues} (2 and 4), \textbf{having to move slowly} (2 and 2), and \textbf{getting no distance information} (3 and 2). 
Furthermore, \lheight{} was reported to suffer from label clutter (3), unpredictable label locations (2), and being bad for summarize tasks (2). 
Another major disadvantage for \lvalue{} was the difficulty of going from label to object (5). 

\subsection{Key Findings}
\textbf{General findings.} Based on the quantitative and qualitative results, we summarize our key findings. 
\textbf{In-view labels} are good for simple identify tasks; however, they do require a full spatial scan by the user.
The main advantages of \textbf{out-of-view labels} are that they achieve high performance in visual comparison tasks, provide a fast overview of all objects, and can give immediate spatial cues to their linked object's location. However, when labeling many out-of-view objects, visual clutter can become a problem. We consider this an important topic for future work.

\noindent \textbf{Task-specific findings.}
(1) \textbf{\langle{} was the overall winner.} \langle{} was one of the best in all three visual search tasks, and achieved the highest level of satisfaction compared to all other labels. (2) \textbf{Out-of-view labels were beneficial in \tcompare{} tasks.} 
All three out-of-view object labels performed better than labels for in-view objects in \tcompare{} tasks, for multiple spatially sparse objects. (3) \textbf{\lvalue{} was not good for \tidentify{} and \tsummarize{}.} \lvalue{} performed the worst in \tidentify{} and \tsummarize{} tasks, which was due to the difficulty to link leader lines to their objects. (4) \textbf{\lsituated{} and \lbound{} had very similar performances.} In all task performance and subjective ratings, there is no significant difference between \lsituated{} and \lbound{}. Finally, (5) \textbf{Data size only affected the performance in \tcompare{}.}

\section{Discussion}

Our study aims to answer how to best place labels in AR to support situated visual search.
Based on our study results with five representative label placing techniques, a short answer to the question is that placing the labels according to the \textbf{angular} direction of the object is overall the best option.
Across all tasks, \langle{} performed as well as or better than other tested techniques.
In addition to identifying the best performing technique, in this section, we discuss finer-grained findings.

\subsection{AR Label Properties and Considerations}

Visual search is a complicated process, which includes multiple components~\cite{laviola20173d, nilsson_natural_2018,lam_framework_2008,yang_embodied_2021}.
To provide a more nuanced understanding of designing AR labels for situated visual search, we discuss the potential explanation and implications of our results with four different components in visual search as compiled by Yang et al~\cite{yang_embodied_2021}:
(a) \textbf{Wayfinding} is the process of finding the destination. 
(b) \textbf{Travel} refers to the motor part of moving to the destination (i.e., walking, rotating). 
(c) \textbf{Context-switching} refers to the extra mental effort that is required to re-interpret a changed view.  
(d) \textbf{Number-of-travels} costs occur when completing a task may involve more than one travel, due to different capabilities of visual representations, or limited working memory.

\noindent
\textbf{Providing labels for out-of-view objects reduces the number of travels in comparison tasks.}
In \tcompare{}, all conditions with out-of-view object labels were faster than the conditions with only in-view object labels.
We conjecture that a potential reason is that providing out-of-view object labels can reduce the number-of-travels 
when comparing multiple spatially sparse targets.
The \tcompare{} task asked participants to search for the object with the highest value.
With out-of-view object labels, all objects' values were accessible in the FoV, so users could first identify the target within the FoV and travel only once.
In contrast, with labels only for in-view objects, participants needed to iterate through the candidates with multiple times of travel 
increasing the participants' memory load to keep track of all candidates' information.
Our finding also aligns with the famous Shneiderman Information Seeking Mantra of \emph{``Overview first, filtering and details on demand''}, which suggests the use of overviews to provide guided search~\cite{shneiderman2003eyes}.
Participants' comments support this hypothesis, as 13 participants complained that labels for only in-view objects required a full-spatial scan.
Although the NASA-TLX physical demand rating was not statistically significant, we can see a trend of higher physical demand with labels for only in-view objects (Fig.~\ref{fig:study-result-ranking}  (a)).
One may argue that automatic filtering can be used to reduce the number of candidates. However, in many cases, coming up with informed criteria for automatic filtering can be challenging due to a wide range of user goals~\cite{brehmerMultiLevelTypologyAbstract2013}. 
Providing an overview of the entire information to guide the visual search is important for those scenarios.

\noindent
\textbf{Precise spatial cues reduce the costs in wayfinding and travel.}
\langle{} was as good or better than other techniques across all tasks.
In situated visual search tasks, the most fundamental interaction for users is to physically turn to face the target object. 
We believe that \langle{} provides the most precise spatial cue among the five tested conditions.
It shows the angular direction of all objects, which allows users to estimate how much and in which direction they need to turn to face the target object.
On the other hand, \lheight{} places labels on the left or right boundary of the FoV to indicate the turning direction, but users cannot estimate the turning angle.
\lvalue{} uses a more subtle cue to indicate the turning direction by placing an arrow on the labels.
As a result, in \lheight{} and \lvalue{}, users need to slowly turn towards the indicated direction and frequently check if the target object is already within their FoV.
Label conditions for only in-view objects did not provide any spatial cues, and users had to fully scan the surrounding space.
In summary, the ranking for spatial cue preciseness is: \langle{} $>$ \lheight{} $>$ \lvalue{} $>$ \lbound{} $=$ \lsituated{}.
This ranking aligns with the good performance of \langle{}.
However, \lheight{} and \lvalue{} performed worse than \lbound{} and \lsituated{}, which is not expected from the ranking. We believe their poor performance is mainly due to proximity factors, which we discuss later in this section.

\noindent
\textbf{Well-designed techniques reduce context-switching costs.}
To utilize the precise spatial cue in \langle{}, participants needed to mentally map the spatial information from the AR FoV to the surrounding physical environment.
According to the VR study by Yang et al.~\cite{yang_embodied_2021}, such operations are likely to introduce a high context-switching cost.
However, their results were not reflected in our study, where we found \langle{} to be the best performing technique.
One potential reason is that, in \langle{}, we used a top-down view to show the spatial distribution of objects with the user's AR FOV always facing up.
We believe this design allows the user to easily interpret the ``overview'' provided by \langle{}, thus reduce the context-switching cost. 
Participants' ranking aligns with our finding, where \langle{} was ranked first by 60\% of participants. 
Participants' comments also support this finding, as 9 participants found \langle{} easy to navigate.

\noindent
\textbf{Label-object proximity is important for spatial search.}
Despite providing a spatial cue, \lvalue{} was the slowest condition for \tidentify{} and \tsummarize{} tasks.
In \lvalue{}, labels were placed in fixed positions on the left side of AR FoV. As a result, they were usually far away from the objects shown within the FoV, resulting in the lowest label-object proximity among all conditions.
We believe that this far distance introduced difficulties for users to follow labels to objects.
Furthermore, with increasing distance, the leader lines for connecting labels and objects become longer and can potentially introduce more line-crossings on the screen, making tracing leader lines even more difficult. 
This is also reflected in the participants' comments, where 5 participants found it difficult to link between objects and labels, and 7 mentioned the line-crossings hindering them from completing the tasks.

\noindent
\textbf{Neighborhood consistency facilitates summarizing values by groups.}
In \tsummarize{}, participants needed to compare the average values to two clusters.
Neighborhood consistency means that when objects are neighbors, their labels should also be neighbors to each other. This property can be shown as the lateral position encoding in Table.~\ref{table:1}. \lsituated{}, \lboundary{} and \langle{} all preserve neighborhood consistency. \lheight{} only encodes lateral positions for in-view objects, and \lvalue{} does not encode lateral positions at all. 
As objects in each target cluster are closer together, a condition with higher neighborhood consistency 
can potentially take less time for summarizing the values of that cluster.
The collected performance data confirmed our analysis, where \lsituated{}, \lbound{}, and \langle{} had a similar performance, and were faster than \lheight{}. \lvalue{} had the worst performance.

\noindent
\textbf{Boundary labels did not demonstrate advantages over situated labels.}
We found a similar performance of \lsituated{} and \lbound{} across all tasks.
In 2D visualizations, boundary labeling is a standard way to reduce visual clutter in many scenarios, and was found to be beneficial~\cite{bekosExternalLabelingTechniques2019}. However, we could not confirm the advantage of boundary labels in our AR study.
One possible explanation is the limited FoV in AR, which makes these two conditions very similar to each other.
We also did not formally control for the visual clutter or occlusions in our study.
Future studies are needed to investigate the effectiveness of these two conditions for other scenarios.

\subsection{Implications for Visual Search with AR Labels}
Based on the above discussion, label designs for only in-view objects are suitable for applications where targets are located in the same direction, requiring less physical movement, such as browsing products on the same shelf. Furthermore, \lboundary{} is a good alternative for \lsituated{} when less visual clutter in the center of the FoV is desired, and tasks require more value comparison, such as comparing the prices of books on the shelf.

Labels for out-of-view objects are suitable for situated visual search when multiple targets of interest are scattered in space. In particular, \langle{} supports location navigation tasks well when precise directional cues are desired, such as searching for coffee shops in the area. As an alternative, \lheight{} provides better guidance when the height information is important for searching targets, such as books on a multiple-layer shelf, paintings hung at various heights in the gallery, and targets located at multiple floors. On the other hand, when knowing the precise locations is not as important as value comparison on the objects, \lvalue{} is a good option for both in-view and out-of-view objects. Use cases include comparing data attributes of dynamically changing targets in real-time, such as players' scores in a live sports game.
Furthermore, combining different label placement strategies in a single application can better support specific tasks. For example, after comparing menu items with \lvalue{} labels, users might switch to \langle{} to locate the targets.

\subsection{Limitations, Generalization and Future Work}
\label{sec:future}
While we have followed the guidance and practice from previous work~\cite{marquardtComparingNonVisualVisual2020,raganSimulationAugmentedReality2009,jungEnsuringSafetyAugmented2018}, simulating AR interactions in VR might not be fully representative of a real-world use scenario. 

\textbf{Visually complex backgrounds.} 
Real-world applications may have more visually complex backgrounds, which are likely to affect the ability to precisely perceive visual information (e.g., color~\cite{whitlock2020graphical}).
To minimize the influence from the background, we suggest using visual cues that do not require precise perception in labels (e.g., we used the unit visualization method~\cite{park2017atom} in our study). 
We believe with effective encoding, people can still easily perceive simple visual information even with complex backgrounds.
Thus, our results can largely transfer to these scenarios.
However, there are cases where one would want to encode multivariate data in labels. 
Further studies are required to investigate the influence of complex backgrounds for these scenarios.
Alternatively, we can show the most critical information in a simple and effective design, like ours, and provide on-demand interactions for users to selectively investigate objects of interest.

\textbf{Real-world objects.} 
Objects in the real world can be more complicated than our controlled environment. In our study, we did not control for objects' sizes and occlusion. Therefore, targets were relatively easy to find, which might have led to ceiling effects in accuracy.
Real-world objects may vary largely in size and shape in some cases. Larger objects with unique shapes may easily attract people's attention, which possibly will influence the performance in identifying the target objects.
We also did not explicitly control the occlusion between objects, as this would have resulted in too many testing factors. We have conducted a post hoc investigation of the trials with occluded targets and did not find a significant difference in performance.
We conjectured that is because participants can easily move in space to find a viewpoint that avoids occlusion.
However, in the case of AR labels, with explicit leader lines to direct the user's attention, such influence can be limited. Further studies are required to verify this expectation with varying difficulty in the real-world settings.


\textbf{Dynamic objects and participants.} 
Another important property for some real-world objects is that they are actively moving (e.g., players on a basketball court).
Labeling dynamic objects requires further design considerations. For example, situated labels should move with the objects; therefore, we need to consider peoples' cognitive capacity to track dynamic visual elements~\cite{heer2007animated}, e.g. , by stabilizing label movement. 

Although we did not limit participants' movement, we observed limited body movements in the study, possibly due to the nature of our tested tasks. In real-world scenarios, we believe users may move more frequently in some scenarios, but they move mostly to get close to one specific target. 
As a result, the user does not need to analyze the entire view all the time. Substantially, our results can still largely apply to these scenarios.

\textbf{Scalability.}
There can be cases that have a larger amount of objects to be labeled. We believe our study has tested representative amounts of objects (e.g., a basketball game has 10 players, and a soccer game has 22 players). Considering the limited size of current AR displays, it is challenging to display a larger amount of labels while still keeping the FoV uncluttered. 
One potential solution is to use focus+context technique to selectively enlarge more important labels, which can be determined by the user's position, facing direction, and object properties. Further study is needed to evaluate the effectiveness of such designs.

In addition to the aforementioned designs and studies, properties of AR labels other than label placement and interaction, such as a label's visual properties~\cite{bellViewManagementVirtual2001}, leader line design~\cite{grassetImagedrivenViewManagement2012,tatzgernHedgehogLabelingView2014} and update frequency~\cite{madsenTemporalCoherenceStrategies2016}, are essential in applying AR labels to real-world applications but are not within the scope of this study.



\section{Conclusion}
We presented a thorough design space exploration of AR labels for out-of-view objects to support situated visual search tasks. Based on our classification of the design space, we compared five AR label conditions in a user study, focusing on the aspect of label placement. Our conditions included \lsituated{} and \textsc{Boundary} for in-view objects, and \lheight{}, \langle{}, and \lvalue{} for out-of-view objects. Our main results suggest that (1) Labels for out-of-view objects are beneficial for compare tasks. (2) Angle-encoded labels showed the best performance overall in supporting situated visual searching and were preferred by participants. (3) Value-encoded labels have strengths in compare tasks but are weak in identify and summarize tasks. 

We hope that our classification and quantitative evaluation of AR out-of-view label design will help researchers to design effective labels and interactions for real-life tasks and applications. 
We envision future AR systems that integrate AR labels into their information retrieval systems, thereby enabling users to perform object-based search or to explore multiple levels of details in data in a truly situated setup.

\section*{Acknowledgments}
This research is supported in part by the National Science Foundation (NSF) under NSF Award Number III-2107328, and the Harvard Physical Sciences and Engineering Accelerator Award. 

\ifCLASSOPTIONcaptionsoff
  \newpage
\fi



\bibliographystyle{IEEEtran}
\bibliography{main}
%

%

\vfill\eject

\begin{IEEEbiography}[{\vspace*{-5mm}\includegraphics[width=1in,height=1.1in,clip,keepaspectratio]{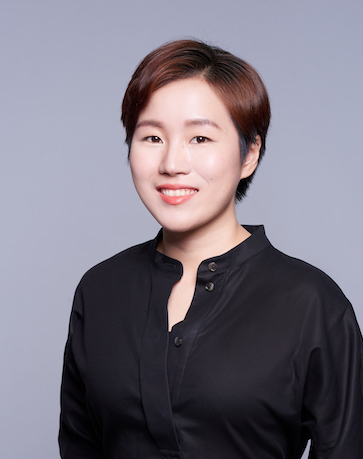}}]{Tica Lin}
is a Ph.D. student at the Visual Computing Group at Harvard University. Prior to joining Harvard, she was a Data Visualization Designer at Visa and a UX Developer at NBA 76ers. Her research interests include data visualization, immersive analytics, and human-computer interaction. In particular, she explores
novel visualization and interaction design in Augmented Reality.
\end{IEEEbiography}
\vspace{-13mm}

\begin{IEEEbiography}[{\vspace*{-5mm}\includegraphics[width=1in,height=1.1in,clip,keepaspectratio]{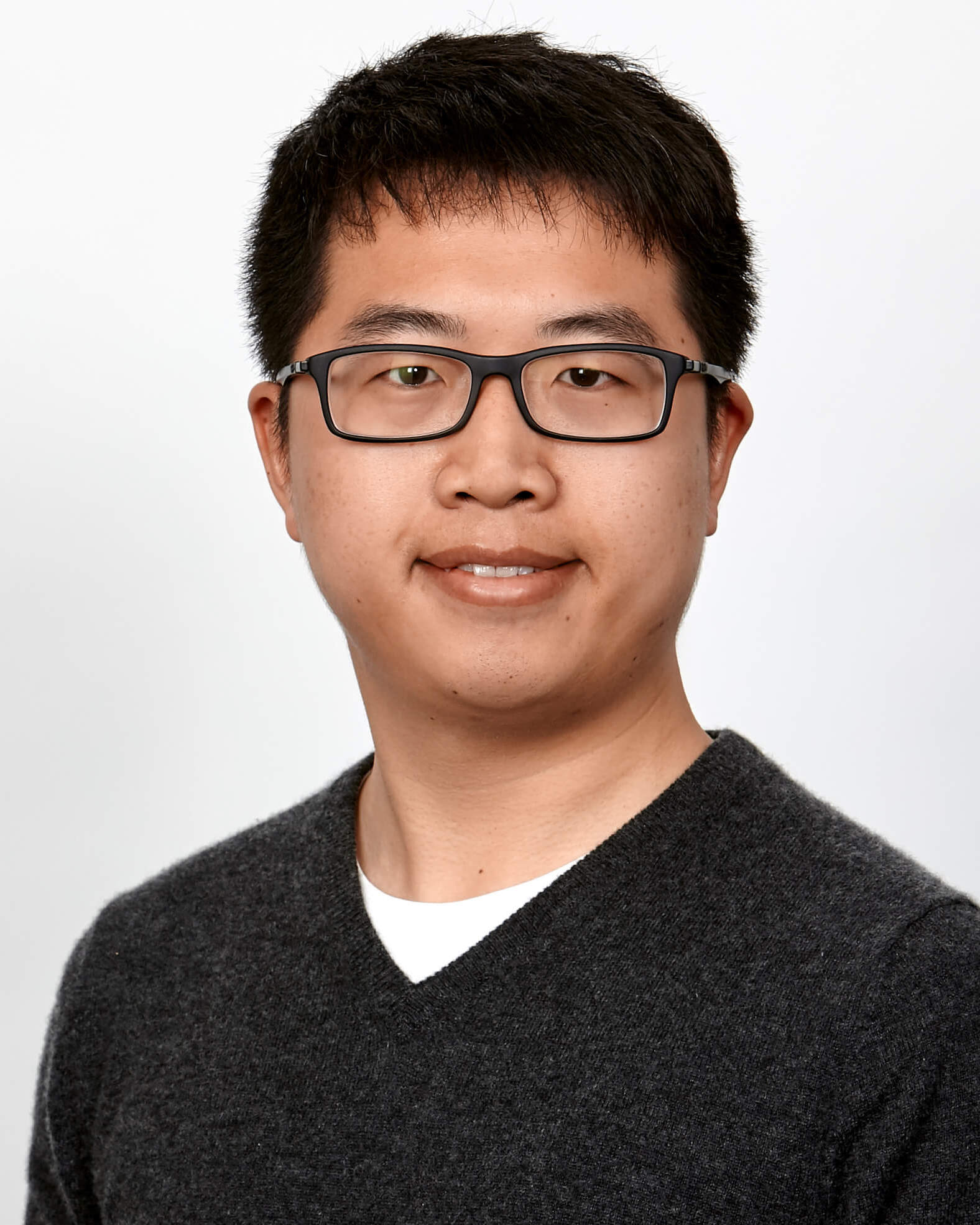}}]{Yalong Yang} is an Assistant Professor at Virginia Tech. He was a Postdoctoral Fellow at the Visual Computing Group at Harvard University, and a Ph.D. student at Monash University, Australia. His research designs and evaluates interactive visualisations on both conventional 2D screens and in 3D immersive environments (VR/AR). He received best paper honorable mentions at VIS 2016 and CHI 2021.
\end{IEEEbiography}
\vspace{-13mm}

\begin{IEEEbiography}[{\vspace*{-4mm}\includegraphics[width=1in,height=1.1in,clip,keepaspectratio]{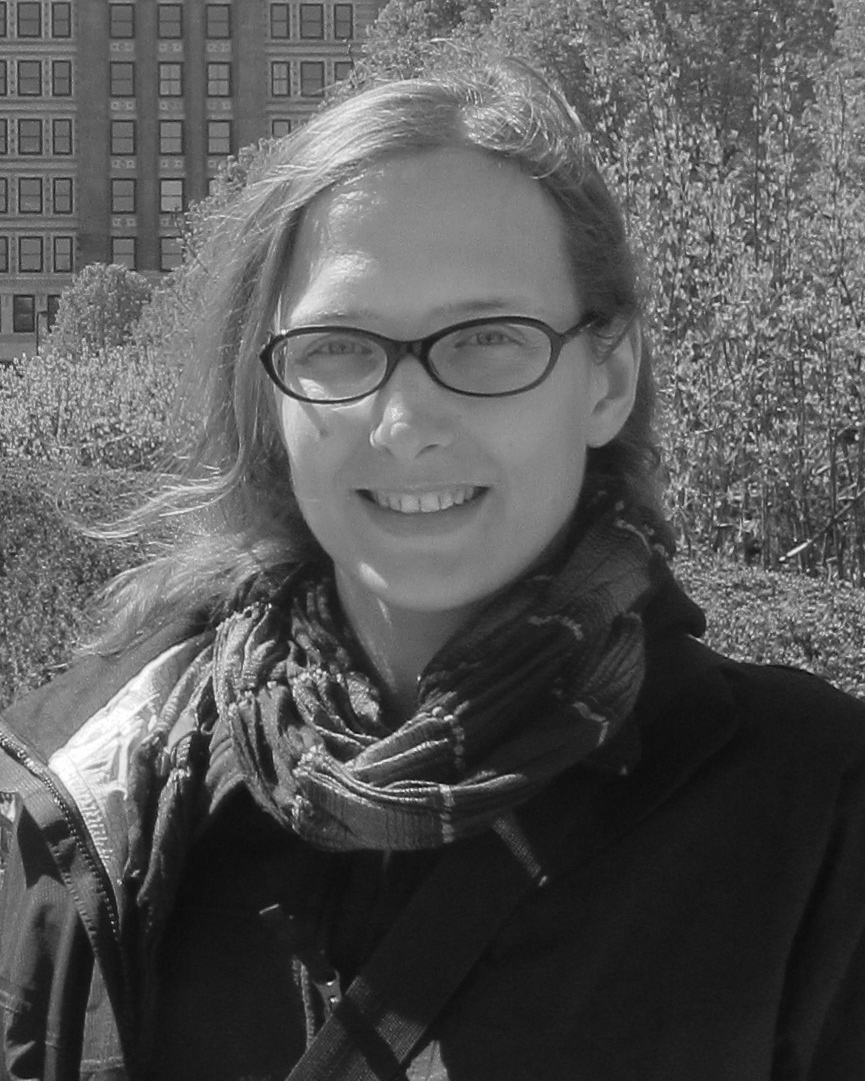}}]{Johanna Beyer}
is a research associate at the Visual Computing Group at Harvard University. Before joining Harvard, she was a postdoctoral fellow at the Visual Computing Center at KAUST. She received her Ph.D. in computer science at the University of Technology Vienna, Austria in 2009. Her research interests include scalable methods for visual abstractions, large-scale volume visualization, and immersive analytics.
\end{IEEEbiography}
\vspace{-13mm}

\begin{IEEEbiography}[{\vspace*{-2mm}\includegraphics[width=1in,height=1.1in,clip,keepaspectratio]{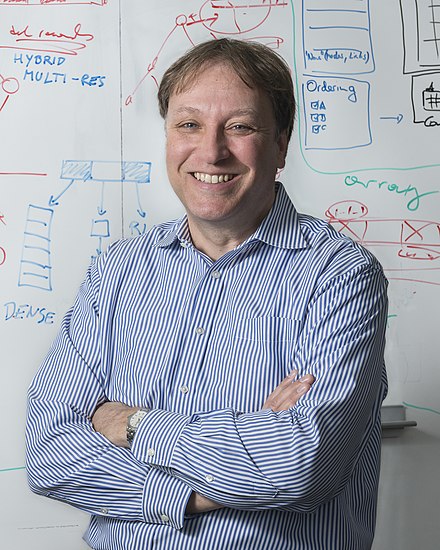}}]{Hanspeter Pfister}
is An Wang Professor of Computer Science in the John A. Paulson School of Engineering and Applied Sciences at Harvard University. His research in visual computing lies at the intersection of scientific visualization, information visualization, computer graphics, and computer vision and spans a wide range of topics, including biomedical image analysis and visualization, image and video analysis, and visual analytics in data science.
\end{IEEEbiography}
\vfill



\end{document}